\documentclass[11pt]{article}
\usepackage{psfrag}
\usepackage{comment,  mathrsfs, bm}
\usepackage{url}
\usepackage{amsfonts, amssymb, latexsym, slashbox}
\usepackage{amsmath}
\usepackage[T1]{fontenc}
\usepackage{tabularx}
\usepackage{comment, url, graphicx, graphics, bm}
\usepackage{natbib}
\usepackage[final]{changes} 
\usepackage{dsfont}
\newlength{\defbaselineskip}
\newcommand{\setlinespacing}[1]%
           {\setlength{\baselineskip}{#1 \defbaselineskip}}

\newcommand{\bernstein}{Bernstein }
\def\o*{o_{{\!}_{P^*}}}
\def\O*{\cal O_{{\!}_{P^*}}}

\def\E{\mathrm{E}}
\def\Var{\mathrm{Var}}

\def\.{\mbox{.}}

\def\sfrac(#1,#2){\mbox{$\frac{#1}{#2}$}}

\def\tr{{\mbox{\tiny{$\mathrm{T}$}}}}

\def\ceiling#1{\left\lceil #1 \right\rceil}

\def\T{{ \mathrm{\scriptscriptstyle T} }}

\newtheorem{theorem}{Theorem}[section]

\newtheorem{remark}{Remark}

\newtheorem{definition}{Definition}

\title
{Efficient and Robust Density Estimation Using \bernstein Type Polynomials}
\author{Zhong Guan\thanks{Email: zguan@iusb.edu
\vspace{6pt}}\\  {\em{
Department of Mathematical Sciences}}\\ {\em{Indiana University South Bend}}\\ {\em{South Bend, IN 46634-7111, USA}}}
\date{}
\begin{document}
\maketitle
\begin{abstract}
  A method of parameterizing and smoothing  the
unknown underlying distributions
 using
 \bernstein type polynomials with positive coefficients is proposed, verified and investigated. Any distribution with bounded and smooth enough density can be approximated by the proposed model which turns out to be a mixture of the
beta distributions, beta$(i+1, m-i+1)$, $i=0,\ldots, m$, for some optimal degree $m$. A simple change-point estimating method for choosing the optimal degree $m$ of the approximate model 
 is presented.
The proposed method  gives a maximum likelihood density estimate which is consistent in
 $L_2$ distance at a
nearly parametric rate \added{${\cal O}(\log n/n)$}  under some conditions.
Simulation study shows that one can benefit from both the smoothness and the efficiency by using the proposed method which can also be used to estimate
some population parameters such as the mean. The proposed methods are applied to three data sets of different types.


\textbf{Keywords:} \bernstein polynomials; Beta Mixture; change-point; Density estimation; Efficiency; Maximum likelihood;   
Model selection;  Nonparametric model; Parametrization; Robustness;   Smoothing.
 
{\em AMS 2000 subject classifications.} Primary  62G05, 62G07, 62G20
\end{abstract}

\setcounter{section}{0}
\section{Introduction}\label{sect: introduction}

Consider the one-sample nonparametric problem in which $x_1,\ldots, x_n$ are independent
observations of random variable $X$ from a population with an unknown
cumulative distribution function (cdf) $F$  and probability density function (pdf) or probability mass function (pmf) $f$. The log-likelihood is
$\ell(f)=\sum_{j=1}^n \log f(x_j)$. If $F$ is continuous or
discrete but with infinite support, then $f$ is an infinite
dimensional parameter \added{in the sense that $f$ belongs to an infinite dimensional space}.
``Since all models are wrong'' \citep{Box-1976-JASA-Sci-and-Stat}, one usually assume the ones which are good approximations to the real world.
\added{For example, in most cases the normal distribution as a parametric model is an approximate model as a result of the central limit theorem.} Because continuous cdf $F$ can always be approximated by nondecreasing step functions, the nonparametric likelihood method
actually assumes that $F$ is a nondecreasing step function with jumps only at the $n$ observations
and uses $p_i=dF(x_i)$, $i=1,\ldots,n$, as parameters. The nonparametric  likelihood is then
defined as $\mathscr{L}_{NP}(\bm{p})=\prod_{i=1}^n p_i$,
where $\bm{p}=(p_1,\ldots, p_n)^\tr$. Then $\bm{p}=(1/n,\ldots,
1/n)^\tr$ is the maximizer of $\mathscr{L}_{NP}(\bm{p})$ subject to
constraints
\begin{equation}\label{basic constraint on p}
 p_i\ge 0,\quad \sum_{i=1}^n p_i=1\.
\end{equation}
So the empirical distribution $\hat F_E(x)=n^{-1}\sum_{i=1}^nI(x_i\le x)$
is the maximum nonparametric likelihood estimate of $F$, where
$I(A)$ is the indicator of $A$. The empirical distribution performs very well as a nonparametric estimate of the underlying cdf $F$ except that it is not smooth when it is supposed to be.
The empirical
likelihood \citep{Owen1988, Owen1990} which makes likelihood ratio based inferences for some
parameters $\theta=\theta(F)$ of interest  such as the population mean 
 also parameterizes the jumps of $F$ at the
observations. This method allows adding estimating equations  which determine $\theta$ and side information
as constraints in addition to (\ref{basic constraint on p}) \added{while maximizing $\mathscr{L}_{NP}(\bm{p})$
to obtain profile likelihood function $\ell(\theta)$ of $\theta$. Then inference can be done based on $\ell(\theta)$}
 \citep[see][ for example.]{QinLawless, owen:2001} \added{As a byproduct the underlying distribution $F$ can be estimated with improved efficiency
 due to useful side information \citep[e.g.][]{Qin-and-Zhang-2005}.}

Nonparametric methods are usually robust but not efficient. Parametric methods are much more efficient but usually lack of robustness. Semiparametric methods balance the trade-off between the efficiency and the robustness but still suffer from model miss-specification.

The nonparametric and the
empirical likelihood methods approximately model the underlying distribution  $F$ by a discrete
distribution  with support $S_F=\{x_1,\ldots, x_n\}$ even if
$F$ is actually continuous.
So  the underlying distribution $F$ is treated as if it were discrete. 
This
is perfectly reasonable only if $F$ is indeed discrete.

Nonparametric density estimation is  more difficult than the estimation of the cumulative distribution function. The
kernel density estimate is commonly used. However it has some drawbacks such as the boundary effect and slow convergence rate. Methods of boundary-effect correction and higher order kernels for improving the performance of the kernel density
estimate have been proposed and studied by many articles \citep[see, for examples,][]{Rice-1984, Devroye-1985-book, Devroye-1989, Devroye-1991}.
 \cite{Grenander-1956-II} introduced the nonparametric maximum likelihood density estimator
in the presence of a monotonicity constraint. 
Grenander's estimator is a histogram estimator. It is non-regular in nature and the usual likelihood properties do not apply.
In particular, \cite{Woodroofe-and-Sun-1993} showed that Grenander's estimator is not consistent at the endpoint.

 In fact, the nonparametric likelihood and the empirical likelihood 
can all be viewed as methods of sieves \citep{Grenander-1981-book, Wong-Shen-1995-Ann-Stat, Shen-1997-Ann-Stat-Sieve} 
in the sense that the
optimizations are carried out within a dense subset of the parameter space.
Consider a certain family 
$\mathscr{D}(S)$ of density functions with support $S$. Let $\mathscr{D}_m(S)$, $m\ge 1$,  be a sequence of dense subfamilies of identifiable parametric density functions such that
each $f_m\in \mathscr{D}_m(S)$ is determined by an $m$-dimensional parameter and for any given sample $x_1,\ldots,x_n$ from $f\in \mathscr{D}(S)$ and each $m$, there is a unique
 $\hat f_m\in \mathscr{D}_m(S)$  that maximizes  $\ell(f_m) = \sum_{j=1}^n\log f_m(x_i)$, i.e.,
 $\ell(\hat f_m)=\max_{f_m\in \mathscr{D}_m(S)}\ell(f_m)$. Note that
maximizing the likelihood $\ell(f_m)$ is equivalent to minimizing the Kullback-Leibler divergence between $f_m$ and the empirical distribution which converges to the true
distribution
as $n\to\infty$. Therefore, if there exists $f_m$ that converges to $f$ fast enough as $m=m(n)\to\infty$ and $n\to\infty$, then we can expect that
$\hat f_m$, as \added{approximate} parametric density estimate,  converges to $f$ at a \deleted{parametric rate or an almost parametric} rate
\added{much faster than that of the nonparametric density estimate such as the kernel density}.

Any continuous (density) function $f$ defined on $[0, 1]$ can be approximated by the \bernstein polynomial \citep{Bernstein}  of degree $m$, 
\begin{equation}\label{m-th order bernstein expansion}
    f_m(t)=\mathbb{B}_{m}f(t)=\sum_{i=0}^{m}
f\big(\sfrac({i},{m})\big)b_{mi}(t),\quad 0\le t\le 1,
\end{equation}
where $b_{mi}(t)={m\choose i}t^i(1-t)^{m-i}$, $i=0,\ldots,m$,
are the \bernstein basis polynomials and $\mathbb{B}_{m}$ is the \bernstein operator.
It is known that the best convergence rate of $ \mathbb{B}_{m}f(t)$ to $f(t)$ is $O(m^{-1})$ if $f$ possesses bounded second or even higher order derivatives.
The \bernstein polynomials  have been used for the purpose of smoothing in statistics by many authors. Most of its applications
are based on the empirical distribution $\hat F_E$.
For example, \cite{Vitale1975}
proposed to estimate $F$ and $f$, respectively, by
$\tilde F_B(t)=\mathbb{B}_{m}\hat F_E(t)
=\sum_{i=0}^{m+1} \hat F_{E}\big(\sfrac({i},{m+1})\big)b_{m+1,i}(t)$ and $\tilde
f_B(t)=\frac{d}{d t}\tilde  F_B(t)$.
%
\cite{Tenbusch1994} extended this method to multidimensional
situations also using empirical distributions. \cite{Petrone-1999-CJS, Petrone-1999-ScadJS} proposed the Bayesian \bernstein density estimates with a \bernstein polynomial prior \citep[see also][]{Petrone-Wasserman-2002-JRSSB, Trippa-etal-2011-AISM}. \added{Methods of bias-corrections for $\tilde F_B$ were discussed by \cite{Babu-2002-JSPI-Bernstein-Ploy-Est},  \cite{Leblanc-2010-JNS},  \cite{Igarashi-and-Kakizawa-2014-JNS} Igarashi and Kakizawa (2014). \cite{Leblanc-2012-aism, Leblanc-2012-JSPI} studied the boundary behavior of the
\bernstein estimator $\tilde F_B$.}
 The rates of convergence of the Bayesian \bernstein density estimate were obtained by
\cite{Ghosal2001}. \cite{Chen-SX-1999-Comput-Stat-Data-Anal-Beta-Kernel-density}
 and \cite{Kakizawa-2004-JNS-Bernstein-density} use \bernstein
polynomials to estimate a probability density by properly choosing different kernel estimates of $f(i/m)$.
\added{The estimation of $F$ using \bernstein polynomials was also studied by many authors
\citep[e.g.][]{Leblanc-2012-aism, Leblanc-2012-JSPI}. Applications of the \bernstein polynomials in spectral density
estimation can be found in
\cite{Kakizawa-2006-J-Time-Ser-Anal}.}
\cite{Guan:Wu:Zhao-2008} applied the \bernstein
polynomial estimate $\tilde f_B$ in the estimation of the proportion of true null hypotheses and the  false discovery rate (FDR).
\bernstein polynomials have been applied to some
regression problems
\citep{Chak-2005-Econmics-letters,
Chang-et-al-2009-IMS-Lecture-Notes,Rafajlowicz-et-al-1999-Nonpar-Bernstein-Regression}.
Studies in the literature also showed advantage of the \bernstein polynomials for shape restricted density and curve estimation \citep[e.g.,][]{Osman-Ghosh-2012-CSDA, Wang-and-Ghosh-2012-CSDA}. As shown by \cite{Babu-2002-JSPI-Bernstein-Ploy-Est}, based on simulation for small and moderate sample sizes, estimator
$\tilde f_B(t)$ may be preferable to the kernel density. However, its convergence rate seems not as attractive as that of the kernel density estimator. Neither
$\tilde f_B(t)$ nor the kernel density $\hat f_K$ is a maximum likelihood estimate.

If $f$ has higher order derivatives, the so-called  iterated \bernstein polynomials $\mathbb{B}_{m}^{(k)}f(t)=\sum_{i=0}^{m}$
$f^{(k)}_{mi}b_{mi}(t)$ provide even better approximations, where the coefficients $f^{(k)}_{mi}$'s  depend on $f(i/m)$, $i=0,1,\ldots,m$, only.
It has been shown that \citep[see, for example,][]{Felbecker-1979, Dzhamalov-1985}
 that if $f$ has bounded $(2k)$th derivative on $[0,1]$, then
$f(t)-\mathbb{B}_{m}^{(k)}f(t)=\mathcal{O}(m^{-k})$.
Similar results can be found in
\cite{Sevy-1993-Calcolo}, \cite{Gonska-Zhou-1994-J-Comp-App-Math}, and  \cite{Adell-etal-1997-J-Math-Anal-Appl}.
Note that $\mathbb{B}_{m}^{(k)}f(t)$ can be written as a linear combination of the density functions $\beta_{mi}(t)\equiv (m+1)b_{mi}(t)$
 of the beta distributions $beta(i+1, m+1-i)$, $i=0,\ldots,m$.
\added{However, the coefficients 
of $\beta_{mi}(t)$ are not necessarily all nonnegative.
Fortunately  as shown by \cite{Lorentz-1963-Math-Annalen}, under some additional conditions, $f$ can be approximated with the same rate as $\mathbb{B}_{m}^{(k)}f(t)$ by a \bernstein\!\!-type polynomial  of the form
 $f_B(t; \bm{p}_m)=  \sum_{i=0}^m   p_{mi}\beta_{mi}(t),$
where $\bm p_m=(p_{m0}, p_{m1}, \ldots, p_{mm})^\T$, and $p_{mj}\ge 0$, which is called a polynomial with  positive  coefficients in the literature of polynomial approximation. The uniqueness of the best approximation was proved by  \cite{Passow-1977-JAT}.
}
Therefore, we can approximately model and parameterize a density $f$ by $f_B(t; \bm{p})$ 
as a mixture of the beta distributions  and
estimate $p_{mi}$ as
parameters using the maximum likelihood method.
Although the model was motivated by density smoothing, it is a new approach to the nonparametric problems because with a better estimate of the underlying population density, one can make a better inference on the population parameters.
%
%
%
The kernel density estimation of $f$ can also be viewed as a mixture of certain types of base density functions \citep[see \S 1.4 of][for example]{McLachlan2000}.
The mixture components of the proposed  model  $f_B(t; \bm{p})$ are specific beta distributions which are free of unknown parameters. Mixture models of unspecified beta distributions has been used, among others, in Bayesian statistics by \cite{Diaconis-Ylvisaker-1985}, in multiple tests by \cite{Parker-Rothenberg-1988-stat-med} and in microarray data analysis by
\cite{Allison-etal-2002}.

Let $S_F$ be the support of the pdf $f$. If $S_F=[a, b]\ne [0, 1]$, then
we can use the linearly transformed data $y_i=(x_i-a)/(b-a)$ in $[0, 1]$ to obtain estimates $\hat G$ and $\hat g$ of the cdf $G$ and the pdf $g$ of $y_i$'s, respectively.
  Then $\hat F(x)=\hat G\{(x-a)/(b-a)\}$ and $\hat f(x)=\hat g\{(x-a)/(b-a)\}/(b-a)$. Otherwise,
if $S_F$ is infinite, 
 we can choose  $[a,b]\supset[x_{(1)}, x_{(n)}]$ as the 
 finite  support of $F$, where $x_{(1)}$ and  $x_{(n)}$ are the minimum and the maximum order statistics respectively.
Specifically, if (i) $S_F=[a, \infty)$, choose $b$ such that
 $1-F(b)=\mathcal{O}(1/\sqrt{n})$  and $b>x_{(n)}$; if (ii)
$S_F=(-\infty, b]$, choose $a$ such that $F(a)=\mathcal{O}(1/\sqrt{n})$ and $a<x_{(1)}$;
  if  (iii) $S_F=(-\infty, \infty)$, choose $a$ and $b$ such that
both $F(a)$ and $1-F(b)$ are of $\mathcal{O}(1/\sqrt{n})$, and $a<x_{(1)}$ and $b>x_{(n)}$.
In case where the support of $F$ is  unknown and $n$ is large,
choose $S_F=[x_{(1)}, x_{(n)}]$.
By the strong law of large numbers for the extreme  order
statistics, $F(x_{(1)})=\mathcal{O}(n^{-1}\log\log n)$ and
 $F(x_{(n)})=1+\mathcal{O}(n^{-1}\log\log n)$, a.s., \citep{Galambos-book-1978}.
For small $n$ and $S_F=(-\infty,\infty)$, for example, assuming that $\E|X|$ is finite,  one can choose $k=\mathcal{O}(\sqrt{n/\log\log n})$, $a<\min\{x_{(1)}, k\}$ and $b>\max\{x_{(n)}, k\}$ so that $P(X<a)=\mathcal{O}(\sqrt{\log\log n/n})$ and $P(X>b)=\mathcal{O}(\sqrt{\log\log n/n})$ by Markov's inequality.
If the support is infinite, using the above linear transformation we actually approximate $f$ by the truncated pdf on $S_F=[a, b]$ with the cdf $F_T(x)=\{F(x)-F(a)\}/\{F(b)-F(a)\}$, $x\in [a,b]$. As it will be shown in this paper that
the best rate of convergence of the estimate $\hat f_T$ of $f_T=f/\{F(b)-F(a)\}$ in the $L_2$ distance is nearly parametric such as
$\mathcal{O}(\log n/n)$. The above chosen interval $[a,b]$ guarantees that  the $L_2$ distance between $f$ and $f_T$ is of $\mathcal{O}(\log n/n)$.
\added{The readers are referred to \cite{Biau-and-Cadre-2008-J-Multivar-Anal} for the exact rate in the density support estimation and the references therein for related results.}

The paper is organized as follows.
In Section \ref{sect: One-sample model} we introduce the \bernstein likelihood based on the special beta mixture model and prove its identifiability. The methods to obtain the corresponding maximum
\bernstein likelihood estimates  of $F$ and $f$ are also described.
 The methods of choosing 
the optimal degree $m$ of the \bernstein polynomial model are presented in Section \ref{sect: choose optimal m}.
Some asymptotic results are given in Section \ref{asymptotic results}.
Simulation results on the performance of the proposed estimates are shown in Section \ref{sect: simulations}. Three illustrative examples are given in Section \ref{sect: examples}.
The proofs of the theorems are relegated
to the Appendix.

\section{Methodology}
\label{sect: One-sample model}
\subsection{\bernstein Polynomial Model}
As having been justified in Section \ref{sect: introduction}
we can approximately parameterize the underlying density function on $[0,1]$ as a mixture of the beta distributions, beta$(i+1, k-i+1)$, $i=0,\ldots,m$, i.e.,
 \begin{equation}\label{Bernstein Approx of f: f(t, p)}
    f_B(t, \bm{p}_m)= \sum_{i=0}^mp_{mi} \beta_{mi}(t)=\beta_{m0}(t)+\bm{p}_m^\tr\bar{\bm\beta}_m(t),\quad 0\le t\le 1,
\end{equation}
where  $\bm{p}_m=(p_{m1},\ldots,p_{mm})^\tr$, 
$\bar{\bm\beta}_m(t)=\{\beta_{m1}(t)-\beta_{m0}(t),\ldots,\beta_{mm}(t)-\beta_{m0}(t)\}^\tr$ and
\begin{equation}\label{eq: constraints}
p_{mi}\ge 0,\quad i=1,\ldots,m;\quad \sum_{i=1}^mp_{mi}\le 1.
\end{equation}
Let $\mathcal{B}_{mi}(t)=I_t(i+1, m-i+1)$ be the regularized incomplete beta function which is also the cdf of
beta($i+1, m+1-i$),   i.e.
$\mathcal{B}_{mi}(t)=\int_0^t\beta_{mi}(u)du$, $i=0,\ldots,m.$
The cdf $F$ can be approximated by
$$F_B(t,\bm{p}_m)=\sum_{i=0}^m p_{mi} \mathcal{B}_{mi}(t)=\mathcal{B}_{m0}(t)+\bm{p}_m^\tr\bar{\bm{\mathcal{B}}}_m(t),\quad 0\le t\le 1,$$
where $\bar{\bm{\mathcal{B}}}_m(t)=\{\mathcal{B}_{m1}(t)-\mathcal{B}_{m0}(t),\ldots, \mathcal{B}_{mm}(t)-\mathcal{B}_{m0}(t)\}^\tr$.
Clearly $F_B$ is a  continuous cdf and $f_B=F'_B$ is the
corresponding pdf.

Although we approximate the underlying density function by a mixture of the beta densities,
it was actually motivated by the \bernstein type polynomials  which was initially invented by Sergej Natanovic Bernstein \citep{Bernstein-1932}, we would rather call the approximate model (\ref{Bernstein Approx of f: f(t, p)}) the {\em \bernstein polynomial model} \added{or the {\em \bernstein density model} as in \cite{Ghosal2001}}.
For a given $m\ge 0$, define the family of \bernstein densities
$$\mathscr{D}_m([0,1])=\Big\{f_B(t, \bm{p}_m)=\beta_{m0}(t)+\bm p_m^\tr\bar{\bm\beta}_{m}(t)\,:\, 
\sum_{i=1}^m p_{mi}\le 1, \; p_{mi}\ge 0,\quad i=0,\ldots,m.\Big\}$$
Then $\cup_{m=1}^\infty \mathscr{D}_m([0,1])$ is dense in $\mathscr{D}([0,1])$, the family of continuous densities on [0,1].
It is clear that
for each $m\ge 1$, model (\ref{Bernstein Approx of f: f(t, p)}) is  identifiable with respect to $\mathscr{D}_m([0,1])$. That is, for $f_B(\cdot, \bm{p}_m), f_B(\cdot, \bm{p}^*_m)\in \mathscr{D}_m([0,1])$, $f_B(t, \bm{p}_m)\equiv f_B(t, \bm{p}^*_m)$ if and only if 
$\bm{p}_m=\bm{p}^*_m$. Moreover, we have the following result whose proof is given in the Appendix.
\begin{theorem}\label{thm: nested model}
For each $m\ge 0$,  model $f_B(t, \bm{p}_m)$ is nested in  model $f_B(t, \bm{p}_{m+r})$, for each $r\ge1$, i.e., $\mathscr{D}_m([0,1])\subset \mathscr{D}_{m+r}([0,1])$.
\end{theorem}
\subsection{The \bernstein Likelihood} 
We define the following \bernstein  
log-likelihood
\begin{equation}\label{eq: Bernstein semiparametric log-likelihood-1}
    \ell_B(\bm{p}_m)=\sum_{j=1}^{n}\log f_B(x_j,\bm{p}_m)=\sum_{j=1}^{n}\log \left\{\beta_{m0}(x_j)+\bm p_m^\tr \bar{\bm\beta}_{m}(x_j)\right\}.
\end{equation}

\begin{definition} The maximizer $\hat{\bm{p}}_m$ of $\ell_B(\bm{p}_m)$ is called the maximum \bernstein likelihood estimate (MBLE) of
$\bm{p}_m$ and  the maximum \bernstein likelihood estimates $\hat f_B(x)=f_B(x, \hat{\bm{p}}_m)$ and $\hat F_B(x)=F_B(x, \hat{\bm{p}}_m)$ of $f(x)$ and
$F(x)$
are called the \bernstein probability density function (BPDF) and the \bernstein  
cumulative distribution function (BCDF), respectively.
\end{definition}
Note that unlike the empirical process $\sqrt{n}\{\hat F_E(t)-F(t)\}$
which is a random
element of the Skorokhod space $D[0, 1]$ of
right-continuous functions on $[0, 1]$ with existing left limit at
each point of $(0, 1]$, the random process $\sqrt{n}\{\hat
F_B(t)-F(t)\}$ is a random element of the space $C[0, 1]$ of
continuous functions on $[0, 1]$.
\subsection{EM and Newton Algorithms}
Since the \bernstein model (\ref{Bernstein Approx of f: f(t, p)}) is actually a finite mixture of $m+1$ completely known beta distributions, the EM algorithm \citep{Dempster1977, Wu1983} can be applied to find the maximum likelihood estimate $\hat{\bm{p}}_m$ of $\bm{p}_m$.
The EM algorithm turns out to be a very  simple iteration \citep[Theorem 4.2 of ][]{Redner-Walker-1984-siam}
\begin{equation}\label{eq: EM iteration for p}
p_{mi}^{(s+1)}=
\frac{1}{n}\sum_{j=1}^{n}\frac{p_{mi}^{(s)}\beta_{mi}(x_j)}{\sum_{k=0}^m p_{mk}^{(s)}\beta_{mk}(x_j)},
\quad i=0,\ldots,m;\; s=0,1,\ldots.
\end{equation}
\cite{Redner-Walker-1984-siam} proved that $\bm p^{(s+1)}_m=(p_{m1}^{(s+1)},\ldots,p_{mm}^{(s+1)})^\tr$ converges to the maximum likelihood estimate.
%
An advantage of the above EM iteration is that any constraints for known zero mixture proportions can be easily imposed by assigning zero initial components of $\bm{p}^{(0)}_m$. Unless it is known to be zero, the initial value of $p_{mi}$ should be nonzero.
The quasi-Newton method using the gradient 
\citep{Broyden-1970-I,Fletcher-1970-Comput-J,Goldfarb-1970-Math-Comput,Shanno-1970-Math-Comp.}
works quite well for searching the maximizer of $\ell_B(\bm{p}_m)$ and is  fast  and  stable.
For each $m$, we can  calculate the beta density iteratively. 
For $t=1$, $\beta_{mi}(1)=0, \quad i=0,\ldots,m-1,\quad \beta_{mm}(1)=m+1.$
For $t\in [0,1)$, $\beta_{m0}(t)=(m+1)(1-t)^m$,
$\beta_{m,i+1}(t)= \{(m-i)t/(i+1)(1-t)\} \beta_{mi}(t)$, $i=1,\ldots,m-1.$

The initial guess of $\bm{p}_m$ can be chosen to be the masses of the discrete uniform distribution over $\{0,1,\ldots,m\}$:
$p_{mi}^{(0)} = {1}/({m+1})$, 
$i=0, \ldots, m\.$
If the sample mean $\bar x$ satisfies $0<\bar x<1$, then we can try
$\bm{p}^{(0)}_m$ as the probability masses of the binomial
distribution $b(m, \bar x)$. Other initials might be $p_{mi}^{(0)}
=\tilde F_B\{(i+1)/(m+1)\}-\tilde F_B\{i/(m+1)\}$, $i=0,\ldots,m$.
Even better initial values seem to be
$\tilde{\bm{p}}_m=
\{I_m(1)\}^{-1} \{\frac{1}{n}\sum_{j=1}^n \bar{\bm\beta}_{m}(x_i)-\int_0^1\beta_{m0}\bar{\bm\beta}_{m}(t)dt \},$
where $I_m(1)$ is given by   (\ref{eq: information matrix})  in  Section \ref{asymptotic results} below.
\subsection{Estimating a functional of $F$}
Let $\theta=\theta(F)$ be a $d$-dimensional parameter and  determined by
$\int \Psi(t,\theta) dF(t)=0$,
where $\Psi(t,\theta)$ is usually a $d$-dimensional vector function.   We can estimate $\theta$ by
$\hat\theta_B=\theta(\hat{F}_B)$.
For example, we can estimate the population mean $\mu=E(X)=\int xdF(x)$ by $\hat\mu_B =\int xd\hat F_B(x)=(m+1)\sum_{i=0}^m \hat p_{mi} \int_0^1
tB_{mi}(t)dt=\sum_{i=0}^m \sfrac({i+1},{m+2})\hat p_{mi}$ with an optimal degree $m$.
The commonly used nonparametric estimator $\tilde\theta=\theta(\hat{F}_E)$ is not always efficient. Because the \bernstein density estimate $\hat f_B$ is an
 approximate maximum likelihood estimate, it is expected that $\hat\theta_B=\theta(\hat{F}_B)$ could gain  efficiency to some degree. It is an interesting research project to study the property of $\hat\theta_B$ and how to construct confidence interval for $\theta$ based on $\hat\theta_B$.

\subsection{Incorporating with Auxiliary Information}
If  auxiliary information is available, more constraints can be added to (\ref{eq: constraints}).
For example, if the boundary values $f(0)$ and $f(1)$ are known, then we have constraints
\begin{equation}\label{eqs: values of density at 0, 1/2, and 1}
f_B(0,\bm{p}_m)=(m+1)p_{m0}=f(0),\quad 
f_B(1,\bm{p}_m)=(m+1)p_{mm}=f(1).
\end{equation}
 If $F$ is symmetric about $1/2$, then $p_{mi}-p_{m,m-i}=0$, $i=0, \ldots, \lfloor m-1 \rfloor/2$,   $m\ge 1$.
  Some inequality constraints such as those resulted from monotonicity of the density \citep{Grenander-1956-II, Robertson-et-al-1988-book} can also be considered.
These constraints can be imposed when using quasi-Newton method in computer package like the constraint optimization in R. The zero
end-point constraints can be added directly in the EM algorithm.

\section{Model Selection}\label{sect: choose optimal m}
The performance of the kernel density estimation depends on  the choice of the kernel and the bandwidth. Although the choice of the kernel plays a less important role, the selection of an optimal bandwidth is difficult because it is a continuous parameter. The Bayesian approach \citep{Petrone-1999-CJS, Petrone-1999-ScadJS} is even more complicated in selecting tuning parameters. The \bernstein polynomial model
is rather specific in the sense that it is determined only by the positive integer $m$.
\subsection{Selecting an Optimal Model Degree $m$}\label{sect: estimating m by chpt method}
The profile Bernstein loglikelihood $\ell(m)=\ell_B(\hat {\bm p}_m)$ is always increasing as $m$ increases. This is similar to many statistical problems in which overfitting can occur. It is interesting project to find an appropriate  penalty term  to create an AIC/BIC-like criterion. In this paper,
we propose to use a change-point detection method to estimate an optimal model degree $m$. Let $M=\{m_0,\ldots,m_k\}$, $m_i=m_0+i$, $i=0,1,\ldots,k$.
We fit the data $x_j$, $j=1,\ldots,n$, with the \bernstein model of degree $m\in M$ to obtain the profile log-likelihood $\ell(m)$. The changes of the log-likelihoods are denoted by
$y_i=\ell(m_{i})-\ell(m_{i-1})$, $i=1,\ldots,k$. Because $\mathscr{D}_m([0,1])\subset \mathscr{D}_{m+1}([0,1])$ for $m\ge 0$
as shown by Theorem \ref{thm: nested model}, we have $y_i\ge 0$, $i=1,\ldots,k$.  We noticed that, as shown in the examples below in Section \ref{sect: examples}, if the optimal degree $m$ is contained in $M$, then
typically, when $m$'s are bigger than the optimal degree, due to over-fitting, the mean and the variation of the corresponding changes are much smaller than those of the changes for $m$'s that are smaller than the optimal degree.
We \replaced{treat }{heuristically  assume that}  $y_1,\ldots,y_\tau$ \replaced{as }{are} exponentials with mean $\mu_1$ and \replaced{treat }{that} $y_{\tau+1},\ldots,y_k$ \replaced{as }{are} exponentials with mean $\mu_0$, where $\mu_1>\mu_0$ and $\tau$ is a change point and $m_\tau$ is the optimal degree. We use the change-point detection method \citep[see  Section 1.4 of][]{Csorgo1997a} for exponential
model to find a change-point estimate $\hat \tau$. Then we estimate the optimal $m$ by $\hat m=m_{\hat \tau}$. Specifically, $\hat\tau=\arg\max_{1\le \tau\le k}\{R(\tau)\}$, where    the
likelihood ratio  of $\tau$ is
$R(\tau)=-\tau\log\{S_\tau/\tau\}-(k-\tau)\log\{(S_k-S_\tau)/(k-\tau)\}+k\log\{S_k/k\}$, $\tau=1,\ldots,k,$ and
 $S_\tau=\sum_{j=1}^\tau y_j=\sum_{j=1}^\tau(\ell_j-\ell_{j-1})=\ell_\tau-\ell_0$, $\tau=1,\ldots,k$. It is obvious that
$$R(\tau)=k\log\left(\frac{\ell_k-\ell_0}{k}\right)-\tau\log\left(\frac{\ell_\tau-\ell_0}{\tau}\right)-(k-\tau)\log\left(\frac{\ell_k-\ell_\tau}{k-\tau}\right),
\quad\tau=1,\ldots,k.$$
If $R(\tau)$ has multiple maximizers, we choose the smallest one as $\hat\tau$.

\added{Although we are not sure about the independence and the exponentiality of  $y_1,\ldots,y_k$, for the purpose of estimating an optimal degree, the above method seems
to work very well as shown by the simulation below in Section \ref{sect: simulations} and examples in Section \ref{sect: examples}. Of course, the theoretical properties of the above method warrant more rigorous
studies. If possible, the set $M$ should be chosen so that both $\tau$ and $k-\tau$ are not small.}
\subsection{Determining the Approximate Lower Bound for $m$ by Mean and Variance}\label{susubsect: determine m by mean and variance}
\added{In order to choose an appropriate starting degree $m_0$ in the above set $M$, we need to know a lower bound for $m$. Let $\mu$   be the mean  of $X$.} If $X$ has density $f(x)\approx f_B(x; \bm p_m)=\sum_{i=0}^mp_{mi}\beta_{mi}(x)$, then it is easy to see
\added{that the variance of $X$ satisfies}
\begin{eqnarray*}
  \sigma^2&=&\Var(X)\approx\frac{m+2}{m+3}\Var(\mu_I)+\frac{\mu(1-\mu)}{m+3}\ge \frac{\mu(1-\mu)}{m+3},
\end{eqnarray*}
where
$\mu_i=\frac{i+1}{m+2}$ is the mean of $beta(i+1, m-i+1)$, and \added{$\Var(\mu_I)$ is the variance of $\mu_I$ as a function of $I$ with distribution}   $P(\mu_I=\mu_i)=p_{mi}$, $i=0,\ldots,m$. 
The equality holds if and only if 
$X$ is $beta(i+1, m-i+1)$ for some $i$.
Therefore we have an approximate lower bound for $m$,
$m_b=\max\left\{1,\ceiling{\mu(1-\mu) /\sigma^{2}-3}\right\},$
 where $\ceiling{x}$ is the ceiling function of $x$.
The special cases of the beta$(a,b)$ with positive integer shape parameters $a$ and $b$  such that $a+b-2\le 1$ are the uniform$(0,1)$=beta$(1,1)$ and the triangle distributions, beta$(1,2)$ and beta$(2,1)$, which have linear densities. The optimal degree for these distributions is $m=1$.
We can estimate   $m_b$ based on the sample data by
\begin{equation}\label{eq: estimator of m}
\hat m_b=\max\left\{\ceiling{{\bar x(1-\bar x)}/{s^{2}}-3}, 1\right\},
\end{equation}
where $\bar x$ and $s^2$ are the sample mean and the sample variance respectively.
The bias of the estimator $\hat\rho={\bar x(1-\bar x)}/{x^{2}}$ of $\rho\equiv\mu(1-\mu)/\sigma^2$ can be reduced
by the jackknife estimate
$\hat\rho_J=n\hat\rho-\frac{n-1}{n}\sum_{i=1}^n\hat\rho_{-i}$,  where $\hat\rho_{-i}$ is the
estimate 
based on leave-one-out sample $\{x_1,\ldots,x_n\}\setminus\{x_i\}$, $i=1,\ldots,n$.
We have done a simulation study which is not shown in this paper in which  
 samples of size $n=20$ were generated from some unimodal distributions. We have compared the empirical or the kernel, and the parametric methods in the estimations of the cdf and the pdf  with the proposed method using  $\hat m_b=\lceil\max\{1,\hat\rho_J-3\}\rceil$ as the model degree $m$ and the jackknife estimate $\hat\rho_J$ of $\rho$.
We have obtained results showing that even with the estimated $\hat m_b$ as the model degree the \bernstein pdf and cdf estimates perform
 similarly to the parametric estimates but much better than the kernel density and the empirical distribution, respectively.

\section{Asymptotic Results}\label{asymptotic results}
The best rate of convergence of the mean integrated squared error (MISE) of the nonparametric kernel density estimate is $\mathcal{O}(n^{-4/5})$ under the assumption that $f$ has
bounded second derivative. This rate is  slower than the $\mathcal{O}(n^{-1})$ convergence rate of the parametric methods.
The kernel density estimate $\hat f_{K}(x)$, the empirical distribution  based \bernstein polynomial estimate $\tilde f_B(t)$, and the maximum \bernstein likelihood density estimate $\hat f_B(t)$ are  similar in the sense that they are all  mixtures of some base densities. However, the beta distributions are not usually chosen in a kernel density estimate. Moreover, the kernel density estimate is not a maximum likelihood estimate.
\cite{Vitale1975}
proposed to estimate   $f$   by
$\tilde f_B(t)$ (see also \cite{Babu-2002-JSPI-Bernstein-Ploy-Est}). Because $\tilde f_B(t)$ is based on the empirical distribution and the \bernstein polynomial, it performs similarly to the kernel density estimation. Except the smoothness, the corresponding cdf $\tilde F_B$ performs similarly
to the empirical distribution $\hat F_E$. \added{We refer the reader to the works of  \cite{Babu-2002-JSPI-Bernstein-Ploy-Est}
and
\cite{Leblanc-2012-aism, Leblanc-2012-JSPI} for details.}

For a subset $\mathcal{I}_m$   of $\{1,\ldots,m\}$, we define
\begin{equation}\label{eq beta bar}
\bar{\bm \beta}_m(t; \mathcal{I}_m)=\{\beta_{mi_1}(t)-\beta_{m0}(t),\ldots,\beta_{mi_\kappa}(t)-\beta_{m0}(t)\}^\tr,
\end{equation}
and the Fisher information matrix for a \bernstein density $f_B$ as 
\begin{equation}\label{eq: information matrix}
I_m(f_B; {\mathcal{I}}_m)
=\int_0^1 \frac{\bar{\bm \beta}_m(t; \mathcal{I}_m)\bar{\bm \beta}_m^\tr(t; \mathcal{I}_m)}{f_B(t)}I\{f_B(t)>0\}dt.
\end{equation}
If $\mathcal{I}_m=\{1,\ldots,m\}$ then $\bar{\bm \beta}_m(t; \mathcal{I}_m)=\bar{\bm \beta}_m(t)$ and $I_m(f_B; {\mathcal{I}}_m)=I_m(f_B)$.

In \cite{Lorentz-1963-Math-Annalen}, $\Lambda_{r,\alpha}=\Lambda_{r,\alpha}(\delta, M_2,\ldots,M_r)$ is defined as the class of
functions $f(t)$ on $[0,1]$ whose first $r$ derivatives $f^{(i)}$, $i=1,\ldots,r$, exist and are continuous  with the properties
\begin{equation}\label{eq: boundary conditions}
    \delta\le f(t)\le M_0,\quad |f^{(i)}(t)|\le M_i,\quad 2\le i\le r,\quad 0\le t\le 1,
\end{equation}
for some $\delta>0$, $M_i>0$, $i=0,2,\ldots,r$, and
\begin{equation}\label{eq: Lipschitz  conditions}
     |f^{(r)}(t)-f^{(r)}(s)|\le |t-s|^\alpha,\quad   0\le s, t\le 1,
\end{equation}
for $0<\alpha \le 1$.

\cite{Lorentz-1963-Math-Annalen} proved that a function $f(t)$ on $[0,1]$ belongs to the class $\Lambda_{r,\alpha}$
if and only if there exists a polynomial $\pi_m(t)$ with positive coefficients of degree $m$ satisfying
\begin{equation}\label{eq: approx of poly w pos coeff}
    |f(t)-\pi_m(t)|\le C \Delta_m^{r+\alpha},\quad 0\le t\le 1,
\end{equation}
where $C=C_r(\delta, M_2,\ldots,M_r)$ depends only on $r$, $\delta$, and the $M_i$, and
$$\Delta_m(t)=\max\left\{\frac{1}{m}, \sqrt{\frac{t(1-t)}{m}}\right\}.$$
It is clear that if $m\le 4$ then $\Delta_m(t)=m^{-1}$, otherwise if $m>4$ then
$$\Delta_m(t)=
\left\{
  \begin{array}{ll}
    \sqrt{{t(1-t)}/{m}}, & \hbox{$|t-0.5|\le 0.5\sqrt{1-{4}/{m}}$;} \\
    m^{-1}, & \hbox{elsewhere.}
  \end{array}
\right.$$
Therefore, if $f\in \Lambda_{2k,\alpha}$ then  $\pi_m(t)$ converges to $f(t)$ at a rate of ${\cal O}(m^{-(r+\alpha)/2})$.

Let $\hat f_B(t)=\beta_{m0}(t)+\hat{\bm{p}}_m^\tr\bar{\bm\beta}_{m}(t)$ be a
maximum likelihood estimate of $f$. Define $\kappa_{mn}^{(r)}=\kappa_{mn}^{(r)}(f)=\sup_{\hat{\bm p}_m}\sum_{i=1}^mI(|\hat p_{mi}-p_{mi}|>n^{-r})$ with $r>(k+1)/2k$.  Clearly $\kappa_{mn}^{(r)}(f)\le m$.
\begin{theorem}
\label{thm: rate of convergence of L2-distance btwn f and fB-hat}
Let $x_1,\ldots,x_n$ be a random sample from a distribution 
with density $f\in \Lambda_{2k,\alpha}$ and
 $C_1n^{1/(2k+1)}\le m\le C_2n^{1/2k}$, for some constants  $C_1$, and $C_2$ which may  depend  on $k$ but are independent of $m$ and $n$. Let $\pi_{m}(t)=\beta_{m0}(t)+\bm{p}_m^\tr\bar{\bm\beta}_{m}(t)$ be the \bernstein density which satisfies
(\ref{eq: approx of poly w pos coeff}) with $r=2k$. then
\begin{itemize}
  \item [(i)]  $\sqrt{n}\{\hat f_B(t)-\pi_{m}(t)\}/\sigma_{mk}(t)$ converges to normal $N(0,1)$ in distribution,
 where
$\sigma_{mk}^2(t)=\Var\{\hat f_B(t)-\pi_{m}(t)\}.$
  \item [(ii)] $\mathrm{MISE}(\hat f_B)=E\int_0^1\{\hat f_B(t)-f(t)\}^2dt = \mathcal{O}(\kappa_{mn}^{(r)} n^{-1})+\mathcal{O}(m^{-2k})$.
\end{itemize}
\end{theorem}
\begin{remark} From (i) of the above theorem, it follows that $\sqrt{n}\{\hat f_B(t)-f(t)\}/\sigma_{mk}(t)$ converges to normal $N(0,1)$ in distribution.
It can also be concluded that $\sqrt{n}\{\hat F_B(t)-F(t)\}$ converges to normal in distribution. Therefore both estimates $\hat f_B$ and $\hat F_B$ are asymptotically
unbiased with biases  equal to zero up to the order $n^{-1/2}$.
The empirical distribution $\hat F_E(t)$ which has the minimum variance among all unbiased estimators of $F(t)$.
However, $\hat F_B(t)$ is a biased estimator  of $F(t)$. So for some $t$,
it could have smaller mean squared error than $\hat F_E(t)$.
Simulation shows that
$\hat F_B(t)$ has smaller  mean squared error than $\hat F_E(t)$ for many distributions.

\end{remark}
\begin{remark}\label{remark about almost parametric rate} Under some additional conditions we have the following conclusions of the assertion (ii) of Theorem \ref{thm: rate of convergence of L2-distance btwn f and fB-hat}.
\added{If $f=\pi_m$ for some $m$, then $f\in \Lambda_{2k,\alpha}$ for all $k>1$. When $k=0.5\log n/\log m$ and $m$ is fixed, $\mathrm{MISE}(\hat f_B)={\cal O}(n^{-1})$. This is the parametric rate of convergence.}
Because $\kappa_{mn}^{(r)}\le m$, if $m=C_1n^{1/(2k+1)}$, then
the rate for $\hat f_B$ is at least as good as ${\cal O}(n^{-2k/(2k+1)})$. 
This is the rate that could be achieved by the nonparametric  kernel density estimate.
If $f\in C^{(\infty)}[0,1]$, then $\mathrm{MISE}(\hat f_B)={\cal O}(n^{-1+\epsilon})$, for all $\epsilon>0$. This is a nearly parametric rate.
 If $\kappa_{mn}^{(r)}={\cal O}(\log m)$ and
$m=C_2n^{1/2k}$, then $\mathrm{MISE}(\hat f_B)={\cal O}(n^{-1}\log n)$. This is also nearly the convergence rate of the parametric density estimate.
\cite{Leblanc-2010-JNS} has proposed a bias-reduction method using the \bernstein polynomial to achieve MISE rate of $\mathcal{O}(n^{-8/9})$.
\end{remark}
\begin{remark}
The rate of convergence in $L_1$ and Hellinger distances of the Bayesian \bernstein density estimate if $f\in \mathscr{D}_m([0,1])$ is shown to be $\mathcal{O}(\log n/\sqrt{n})$ by \cite{Ghosal2001}. If the true density does not belong to $\mathscr{D}_m([0,1])$, \cite{Ghosal2001} also showed that the best rate achieved by the Bayesian \bernstein density estimate  is $\mathcal{O}(n^{-2/5})$.
\end{remark}
\begin{remark}
By a theorem of \cite{Jackson-1930} (see also \cite{de-la-Vallee-Poussin-1970} or pages 19--20 of \cite{Lorentz-1986-book-bernstein-poly}), we know that
if $f\in C^{(k)}[0,1]$, then for each $m>k$ there is a polynomial $P_m(t)$ of degree $m$ such that $|f(t)-P_m(t)|={\cal O}(m^{-k})$.
Let $\tilde\pi_m(t)$ be the \bernstein density  which satisfies
(\ref{eq: approx of poly w pos coeff}) with $r=2k$ and $f=P_m$. Therefore  we have
$|f(t)-\tilde\pi_m(t)|\le |f(t)-P_m(t)|+|P_m(t)-\tilde\pi_m(t)|=\mathcal{O}(m^{-k})$.
This means that the condition $f\in \Lambda_{2k,\alpha}$ can be reduced to $f\in \Lambda_{k,\alpha}$.
\end{remark}

\section{Simulation Study}\label{sect: simulations}
We have conducted simulation using the estimated optimal degree $\hat m$ of Section \ref{sect: estimating m by chpt method} for  sample sizes $n=20,   100$, and  $500$ based on 500 Monte Carlo runs with samples generated from the following distributions. We used the R function \textsf{density()} for calculating $\hat f_K$ using normal kernel and the commonly recommended  method of \cite{Sheather-and-Jones-1991-JRSSB} to choose the bandwidth.
\subsection{Distributions Used in the Simulation}
\begin{itemize}
    \item [(i)]  B$(a,b)$: the {beta} distribution with
 $\mu=a/(a+b)$ and $\sigma^2=\mu(1-\mu)/(a+b+1)$. 
  So
 $m_b
 =\max\left\{1,\ceiling{a+b-2}\right\}.$
    \item [(ii)] G($\alpha,\beta$): the {gamma} distribution  with mean $\mu=\alpha\beta$ and variance $\sigma^2=\alpha\beta^2$. We truncate this distribution by the  interval $[0, \mu+k\sigma]$ with $k=5$. Thus
        we have $m_b=5$ for gamma(2,2).
    \item [(iii)] N($0, 1$): the standard {normal} distribution  truncated   by the interval $[-5, 5]$. 
    Thus $m_b=23$.
    \item [(iv)] NM: the {normal mixture},  $0.5N(-1,.5^2)+0.5N(1,.3^2)$, truncated by $[-3.5, 3.5]$. For nonunimodal distributions, $m_b$ is usually
    much smaller than the optimal $m$.
    \item [(v)] NN($k$): the {nearly normal}
      distribution  of $\bar u_k=(u_1+\cdots+u_{k})/k$ with $u_1,\ldots,u_{k}$ being independent uniform(0,1) random variables. The density of $\bar u_k$ is denoted by
      $\psi_k(t)$. The lower bound is $m_b=
      3(k-1)$.
    \item [(vi)] NNM: the {nearly normal mixure}
      distribution with density $0.5 \psi_k(x/1.5)/1.5+0.5 \psi_k[(x-1)/2]/2$, $0\le x\le 3$, $k=4$. For this mixture distribution $m_b$ is much smaller than the optimal $m$.
\end{itemize}
It should be noted that the density of NN($k$) satisfies $\psi_k(t)\in C^{(k-2)}[0,1]$ but $\psi_k(t)\notin C^{(k-1)}[0,1]$ for $k\ge 2$. In fact, when, $k\ge 2$,  $\psi_k(t)$ is a piecewise  polynomial function of degree $(k-1)$ defined on pieces $[i/k, (i+1)/k)$, $i=0,1,\ldots,k-1$. In the simulation, we used the normal distributions as the parametric models of NN($4$). For the nearly normal (mixture)
distribution, both the normal (mixture) and the \bernstein model are approximate parametric models. Some of the above distributions are bimodal and some have infinite support. For parametric method, the parameters were estimated using the maximum parametric likelihood method.
\subsection{The Performance of the Estimated Optimal Degree}
 The estimated mean and variance of the optimal degree estimate $\hat m$ are reported in Table \ref{tbl1: simulation using optimal degree m}.  The performance of the proposed method for estimating the optimal degree $m$ is satisfactory. For beta(1,1)  i.e., uniform(0,1), the true optimal degree is $m=0$.
 Based on random samples from this distribution, the empirical optimal degrees as  change-points of the increments of the loglikelihood are usually
 close to 0. This makes it difficult to be detected using the method of change-point estimation.  For the beta distribution with integer parameters $a=5$ and $b=7$, the \bernstein model is a true parametric model with $m=a+b-2=10$. From the simulation results for the beta distribution B(5,7) we see that the change-point method estimate of the optimal $m$ is consistent. The  results for  B(2.5,10) and NN($4$) also indicate that the proposed estimate $\hat m$ selects a satisfactory model degree. The performance of the density and distribution function estimates for all the cases also convinces that $\hat m$  is a good estimator of the optimal model degree $m$.
From this simulation, we also notice that as the sample size increases the optimal model degree $m$ either remains unchanged as for B(5,7) or increases very slowly as commented in Remark \ref{remark about almost parametric rate}.
%
\subsection{The Point-wise  Mean Squared Errors and the Mean Integrated  Squared Errors  for the CDF and the PDF Estimates}
\begin{figure}
\begin{center}
\includegraphics[width=5.5in]{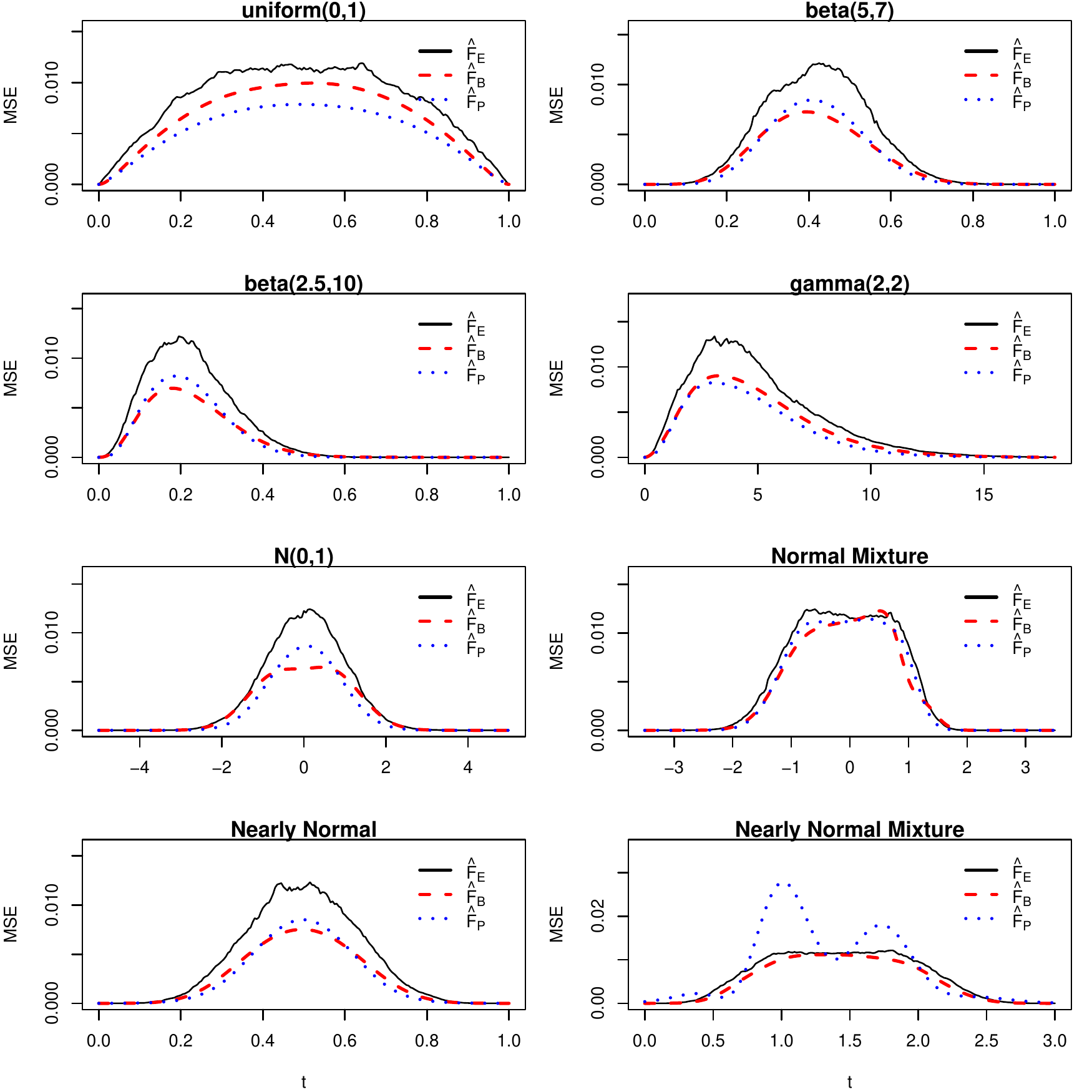}
\caption{\label{fig: pMSES of estimates of cdf based on 500 runs n=20}The point-wise  mean squared errors of the
empirical distribution $\hat F_E$, the parametric estimate $\hat F_P$ and the
maximum \bernstein likelihood estimate $\hat F_B$ with $n=20$ based on $500$ runs.
}
\end{center}
\end{figure}

\begin{figure}
\begin{center}
  \includegraphics[width=5.5in]{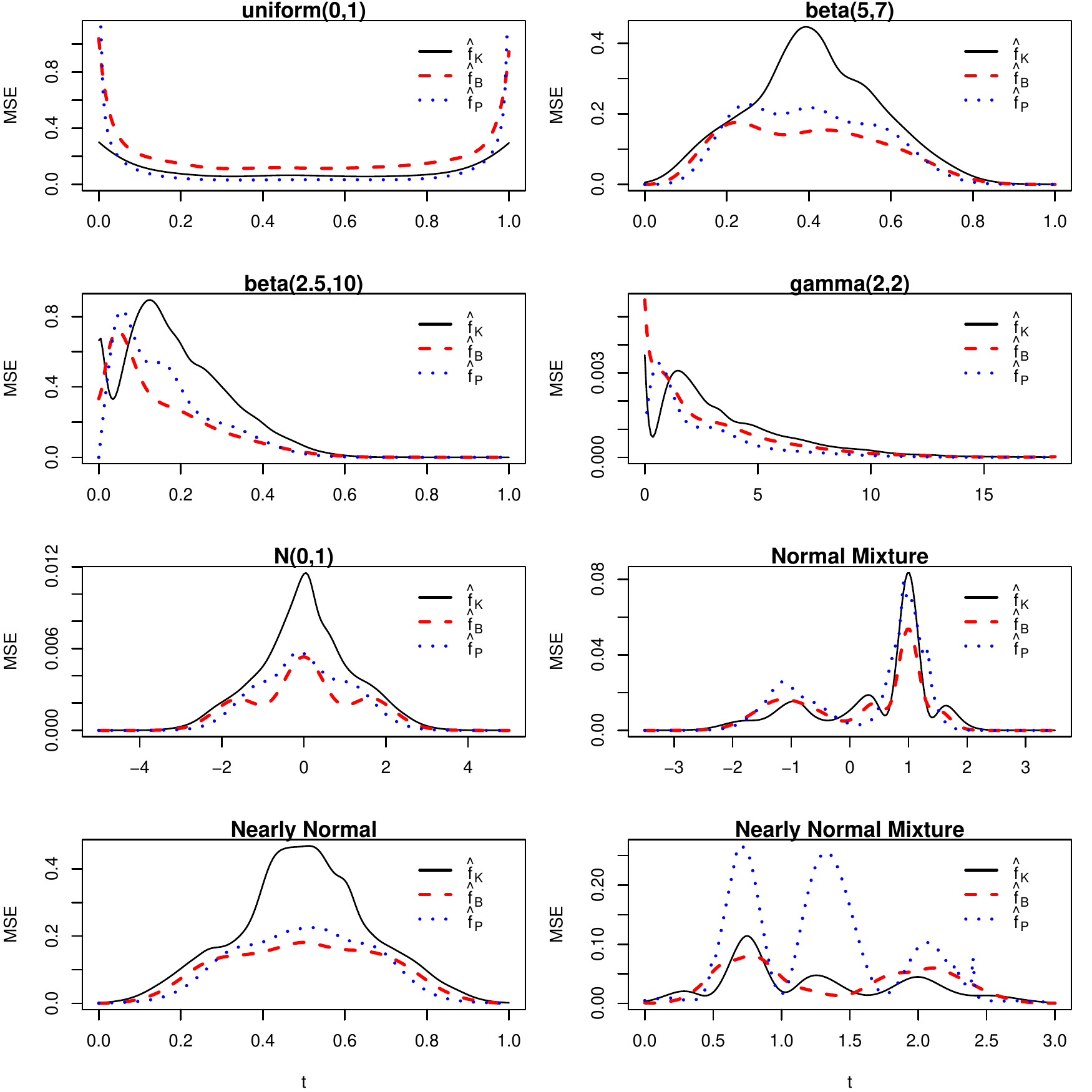}
\caption{\label{fig: pMSES of estimates of pdf based on 500 runs n=20}The point-wise  mean squared errors of the
 kernel density estimate $\hat f_K$, the
parametric density estimate $\hat f_P$, and the
maximum \bernstein likelihood estimate  $\hat f_B$  with $n=20$ based on $500$ runs.
}
\end{center}
\end{figure}

\begin{figure}
\begin{center}
\includegraphics[width=5.5in]{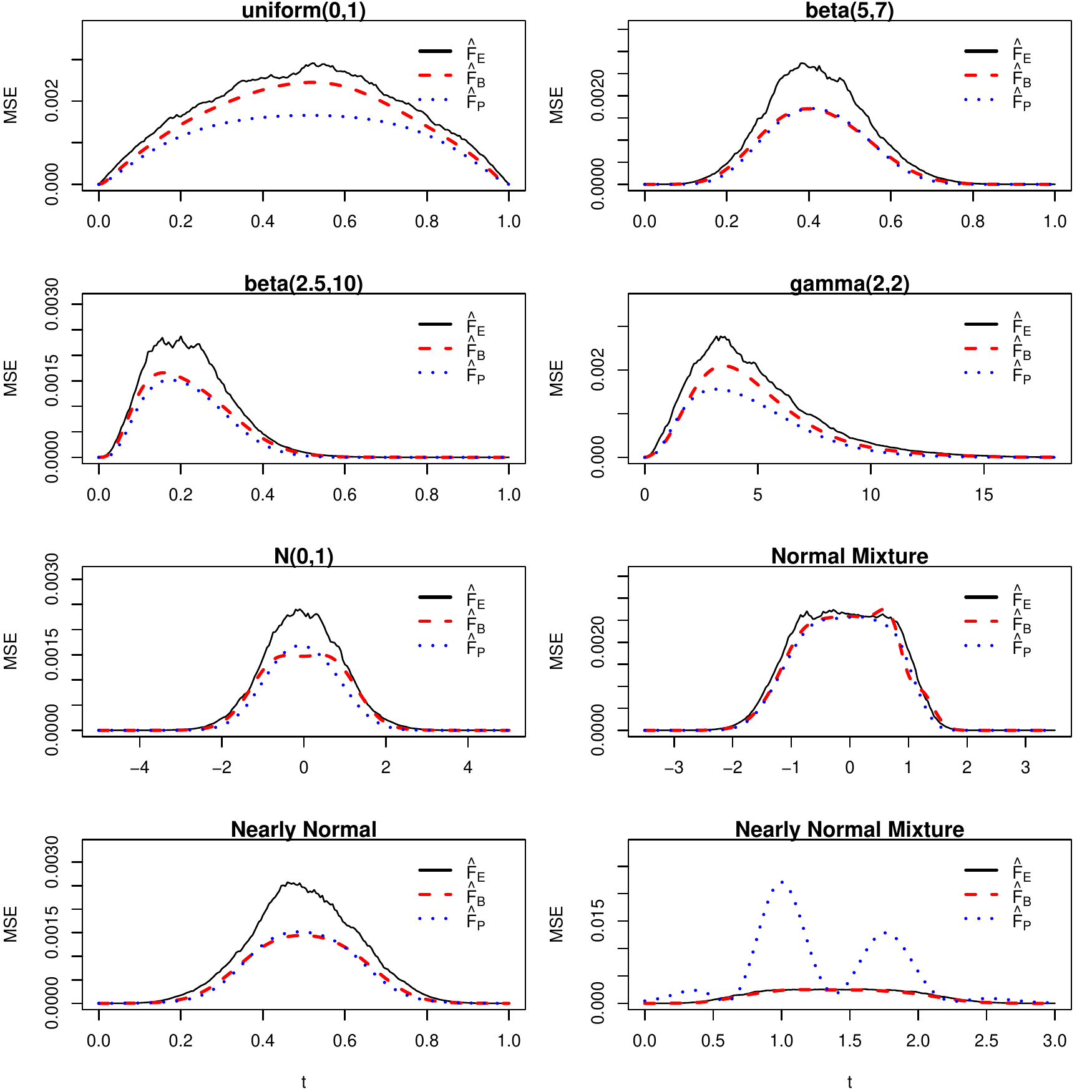}
\caption{\label{fig: pMSES of estimates of cdf based on 500 runs n=100} The point-wise  mean squared errors of the
empirical distribution $\hat F_E$, the parametric estimate $\hat F_P$ and the
maximum \bernstein likelihood estimate $\hat F_B$ with $n=100$ based on $500$ runs.
}
\end{center}
\end{figure}

\begin{figure}
\begin{center}
  \includegraphics[width=5.5in]{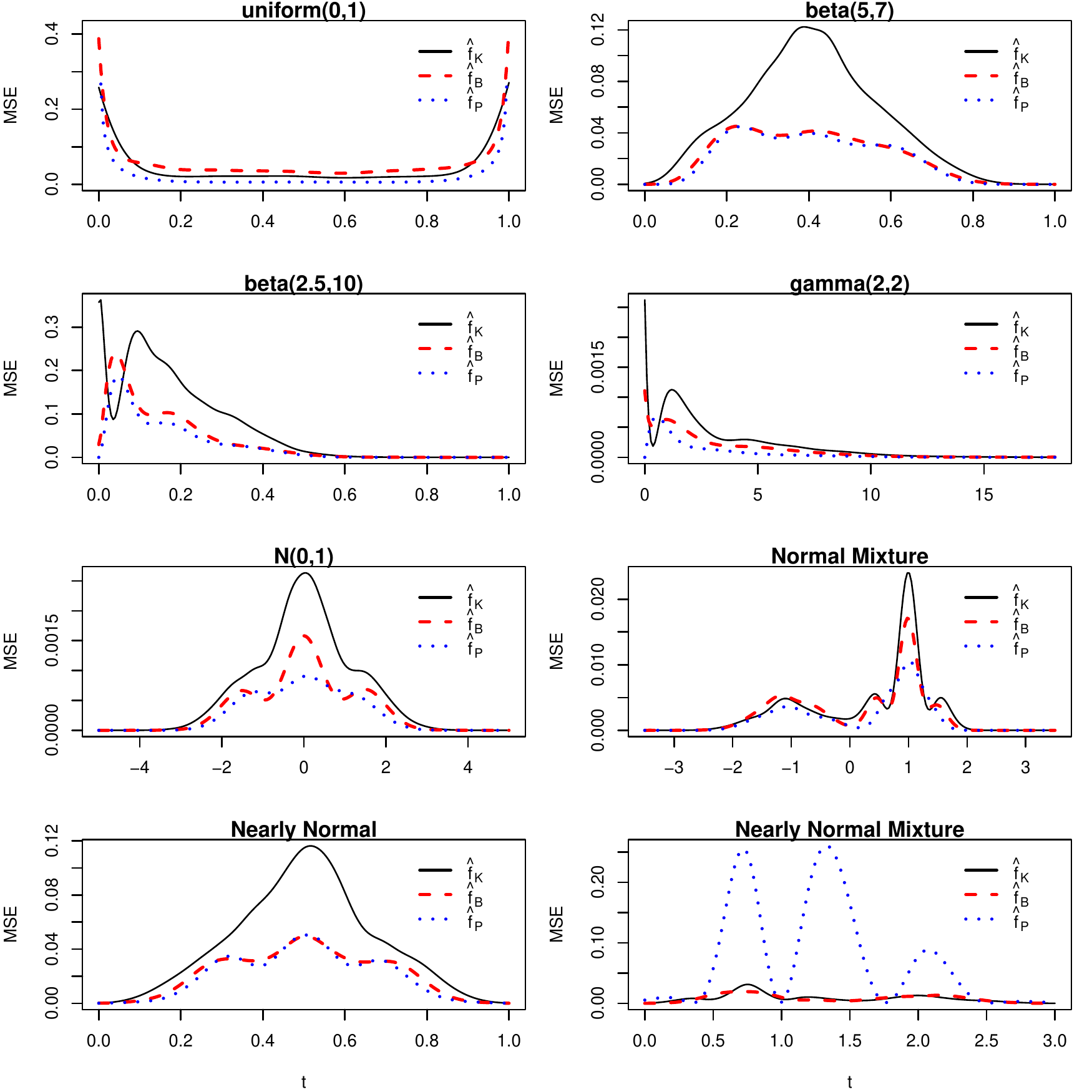}
\caption{\label{fig: pMSES of estimates of pdf based on 500 runs n=100} The point-wise  mean squared errors of the
 kernel density estimate $\hat f_K$, the
parametric density estimate $\hat f_P$, and the
maximum \bernstein likelihood estimate  $\hat f_B$  with $n=100$ based on $500$ runs.
}
\end{center}
\end{figure}

We compared the point-wise mean squared errors of the pdf estimate,
$\mathrm{MSE}(\hat f(t))=E\{\hat f(t)-f(t)\}^2$, for the parametric estimate $\hat f_P$,  the proposed estimate $\hat f_B$ and the kernel density $\hat f_K$
for the above distributions.
We also compared the point-wise mean squared errors of the cdf estimate,
$\mathrm{MSE}(\hat F(t))=E\{\hat F(t)-F(t)\}^2$, for the parametric estimate $\hat F_P$,  the proposed estimate $\hat F_B$ and the empirical distribution $\hat F_E$.
The point-wise mean squared errors were calculated at 200 equally-spaced points on the (truncated) support interval.
From the simulation results shown by Figures \ref{fig: pMSES of estimates of cdf based on 500 runs n=20} through \ref{fig: pMSES of estimates of pdf based on 500 runs n=100} we see that the performance of the proposed pdf(cdf) estimate is between the kernel density (the empirical distribution) and the parametric estimate.  Except for the uniform distribution, in all the other cases, it leans to  the parametric estimate and always much better than the kernel density (the empirical distribution).


The mean integrated squared errors of the density estimates,
$\mathrm{MISE}(\hat f)=E\int_a^b\{\hat f(t)-f(t)\}^2dt$, were calculated for the parametric estimate $\hat f=\hat f_P$,  the proposed estimate $\hat f=\hat f_B$ and the kernel density $\hat f=\hat f_K$. For fast simulation, we  divided the (truncated) support interval $[a, b]$ into 200 subintervals of equal-length for approximating the integral.

From the simulation results shown by Table \ref{tbl1: simulation using optimal degree m} we see that the performance of the proposed density estimate is very similar to the parametric estimate. \added{When samples are from normal mixture distribution, the kernel density with the normal kernel is close to the parametric and the \bernstein densities because it is also some kind of normal mixture model.}  In most cases, it leans to  the parametric estimate and always much better than the kernel density. \added{The \bernstein
method is even  better than the parametric one in some cases especially when samples are from the nearly normal (mixture) distribution. So the proposed method is
robust and efficient.}

\begin{table}
  \caption{The mean and variance of the estimated optimal model degree $\hat m$, the mean integrated squared errors  ($\times 100$) of the density estimates, and the mean squared errors
($\times 100$)  of the estimates of $\mu$ based 500 Monte Carlo runs. $\hat f_P$: the parametric density estimate; $\hat f_B$: the maximum \bernstein likelihood density estimate; $\hat f_K$: the kernel density. Samples were generated from the following distributions.  B($a,b$): beta distributions with $(a,b)=(1,1), (5,7), (2.5,10)$; G(2,2): the gamma(2,2) truncated by $[0, 4+5\sqrt{8}]$;
N(0,1): the standard normal truncated by $[-5,5]$;   NM: normal mixture $0.5N(-1,.5^2)+0.5N(1,.3^2)$ truncated by $[-3.5, 3.5]$;
NN($k$): the distribution of the mean of the independent uniform(0,1) random variables $u_1,\ldots,u_k$; and NNM: the nearly normal mixture.}
{\footnotesize{
\begin{tabular}{*{9}{c}}
\hline
&$\E(\hat m)$&$\Var(\hat m)$&MISE($\hat f_P$)&MISE($\hat f_B$)&MISE($\hat f_K$)&MSE($\hat\mu_{P}$)&MSE($\hat\mu_{B}$)&MSE($\bar x$)\\\hline
&\multicolumn{7}{c}{$n=20$}\\
U$(0,1)$
&~~7.58&~34.50&~8.2955&18.3113&~9.5223&~0.3685&~0.3806&~0.3825\\
B$(5,7)$
&~10.07&~11.95&11.0209&{\em ~8.5695}&16.5928&~0.0861&{\em ~0.0839}&~0.0866\\
B$(\frac{5}{2},10)$
&~11.07&~13.37&15.5951&{\em 12.6858}&24.4440&~0.0637&~0.0650&~0.0635\\
G$(2,2)$
&~11.54&~34.10&~0.0423&~0.0613&~0.0659&36.0210&34.9266&36.0210\\
N$(0,1)$
&~17.79&~17.47&~0.1434&{\em ~0.1262}&~0.2558&~4.9798&~5.0158&~4.9798\\
NM
&~71.20&537.00&~1.0491&{\em ~0.7923}&~0.9405&~5.7849&{\em ~5.7448}&~5.7849\\
NN$(4)$
&~10.23&~12.77&10.3995&{\em ~9.2777}&16.7884&~0.1065&~0.1062&~0.1065\\
NNM
&~47.45&537.58&~7.8040&~3.2116&~2.7099&~2.0042&~1.9941&~2.0042\\
&\multicolumn{7}{c}{$n=100$}\\
U$(0,1)$
&~~8.26&~40.70&~1.7510&~5.2368&~4.4700&~0.0762&~0.0822&~0.0832\\
B$(5,7)$
&~10.56&~~4.63&~1.9919&~2.2092&~4.4836&~0.0193&~0.0203&~0.0193\\
B$(\frac{5}{2},10)$
&~11.11&~~6.07&~2.9263&~3.9513&~7.0469&~0.0127&~0.0141&~0.0128\\
G$(2,2)$
&~~9.40&~10.04&~0.0085&~0.0139&~0.0224&~8.5059&~8.5197&~8.5059\\
N$(0,1)$
&~20.56&~11.02&~0.0272&~0.0364&~0.0657&~1.0192&{\em ~0.9520}&~1.0192\\
NM
&~87.06&294.10&~0.1663&~0.2281&~0.2759&~1.1285&~1.1228&~1.1285\\
NN$(4)$
&~~9.93&~~5.27&~1.8868&~2.2209&~4.0577&~0.0191&~0.0181&~0.0191\\
NNM
&~37.14&223.84&~7.0935&{\em ~0.8678}&~0.8601&~0.4848&~0.4840&~0.4848\\
&\multicolumn{7}{c}{$n=500$}\\
U$(0,1)$
&~~8.58&~43.51&~0.3582&~1.2855&~2.2784&~0.0144&~0.0155&~0.0156\\
B$(5,7)$
&~10.44&~~1.04&~0.3671&{\em ~0.3579}&~1.2796&~0.0038&~0.0033&~0.0038\\
B$(\frac{5}{2},10)$
&~12.14&~~2.44&~0.5897&~1.0097&~2.0994&~0.0025&~0.0025&~0.0025\\
G$(2,2)$
&~~9.14&~~2.48&~0.0016&~0.0028&~0.0073&~1.6251&~1.6035&~1.6251\\
N$(0,1)$
&~22.26&~~4.08&~0.0050&~0.0086&~0.0184&~0.1971&{\em ~0.1685}&~0.1971\\
NM
&106.74&129.07&~0.0329&~0.0562&~0.0762&~0.2447&~0.2449&~0.2447\\
NN$(4)$
&~~9.91&~~2.11&~0.5742&{\em ~0.5848}&~1.2991&~0.0042&{\em ~0.0038}&~0.0042\\
NNM
&~34.20&154.66&~6.9376&{\em ~0.6623}&~0.2363&~0.0940&~0.0931&~0.0940\\
\hline
\end{tabular}}}
\label{tbl1: simulation using optimal degree m}
\end{table}
\subsubsection{\bf The Mean Squared Errors of Estimates of the Population Mean}
To show that the proposed method can result in not only a better density estimate than the kernel density but also estimates of some population parameters nearly as good as the parametric maximum likelihood estimates, we did a simulation to compare the mean squared error $\mathrm{MSE}(\hat\mu_B)$ of the maximum \bernstein likelihood estimate $\hat\mu_B$ of the population mean $\mu$ with those of the parametric maximum likelihood estimate   $\hat\mu_{P}$. 
Of course, in many cases, $\hat\mu_{P}=\bar x$. The simulation results presented in Table \ref{tbl1: simulation using optimal degree m} show that the maximum \bernstein likelihood estimator $\hat\mu_B$ has mean squared errors very close to those of the maximum parametric likelihood estimators.

In Table \ref{tbl1: simulation using optimal degree m}, for distribution B(5,7) which is cast into two different parametric models, both $\hat\mu_P$ and $\hat\mu_B$ are the maximum  parametric likelihood estimates. It seems that the two are equally good. For the nearly normal distribution NN($4$),   the  estimator $\hat\mu_B$ has a smaller mean squared error. In this case both the normal  and the \bernstein models are {\em approximate} models. This shows that the proposed method is robust and efficient.   \added{When samples are   exactly normal $\hat\mu_B$ has even smaller mean squared error for large sample size.}  The normal model is so popular mainly because of the central limit theorem. In many cases, when we use the normal distribution model, we may just have a sample looked like normal but actually from a nearly normal population.

\section{Examples}\label{sect: examples}
In this section, we shall apply our method to three quite different types of data sets. Their distributions are unimodal and slightly skewed, bimodal, and unimodal but extremely skewed, respectively. The optimal model degrees were estimated by the method of change-point.
\subsection{Annual Flow Data of Vaal River}
\begin{figure}
\begin{center}
  \makebox{\includegraphics[width=4.5in]{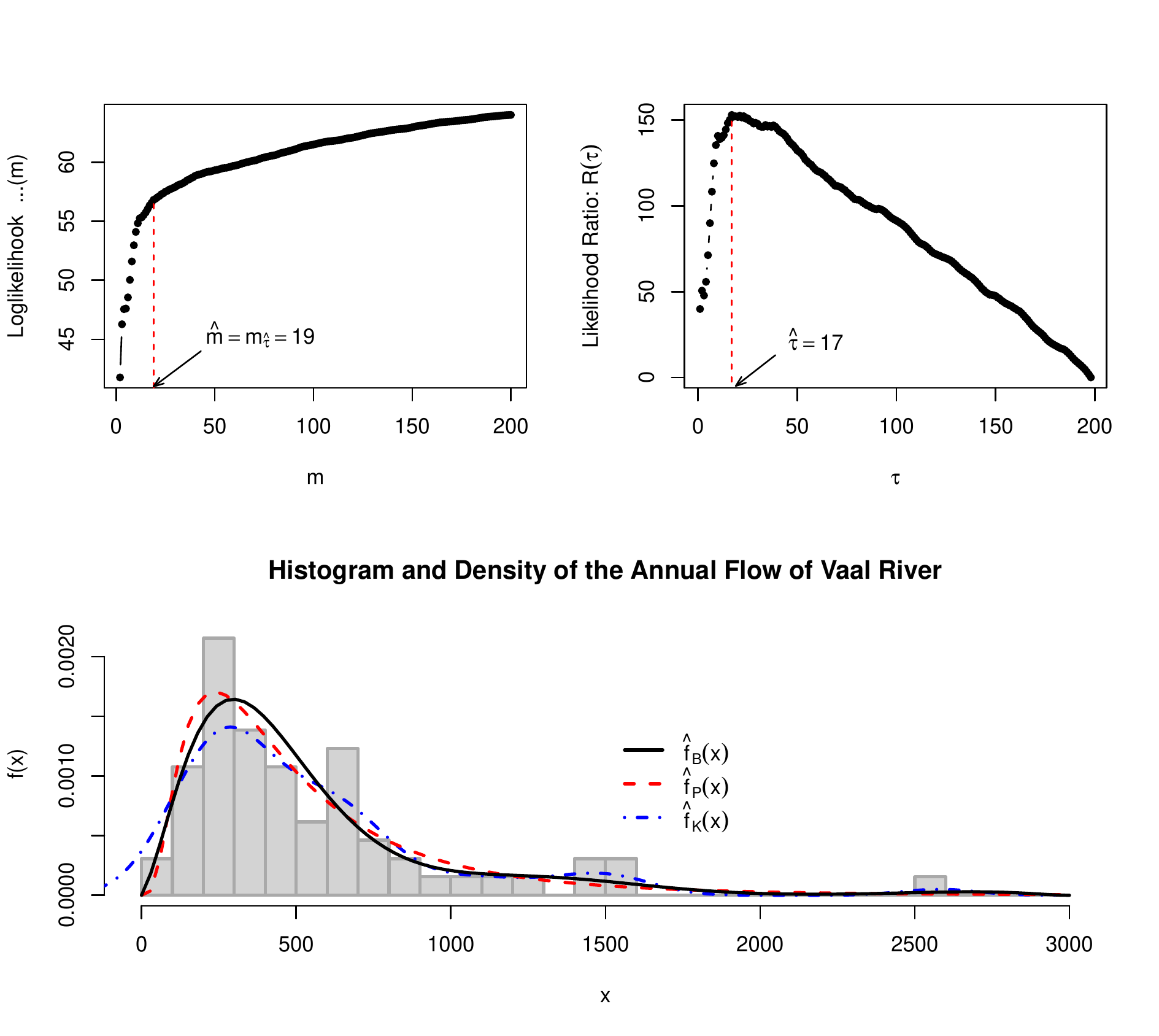}}
\caption{\label{fig: estimated-pdfs-vaal-river-annual-flow-data}
Upper panels: profile loglikelihood $\ell(m)$  and the likelihood ratio $R(\tau)$ for change-point $\tau$ for $m\in M\{2,3,\ldots,200\}$.
Lower panel:  $\hat f_P$:
the log-normal parametric density estimate; $\hat f_K$: the kernel density estimate;  and $\hat f_B$: the proposed maximum  \bernstein likelihood estimate using
 $\hat m=m_{\hat \tau}= 19$.}
 \end{center}
\end{figure}

The annual flow data of Vaal River at Standerton as given by Table 1.1 of \cite{Linhart-and-Zucchini-model-selection-book}
give the flow in millions of cubic meters. The lognormal distribution is compared by these authors with the normal and the gamma distributions and is shown to fit the data better.
Using the \bernstein density model and truncate the data by interval $[a,b]=[0,3000]$, we obtain $\hat m_b=9$.
We obtained the estimated optimal model degree $\hat m=19$ using the change-point method.
Figure \ref{fig: estimated-pdfs-vaal-river-annual-flow-data} shows the estimated density using the lognormal model $\hat f_P$, the
\bernstein density estimate $\hat f_B$, and the kernel density estimate $\hat f_K$. We can see that the \bernstein density estimate is very close to the lognormal density estimate.
%
\subsection{Old Faithful Eruption Duration}
As a bimodal data we consider the density estimation of the duration (in minutes) of eruptions of the Old Faithful based on
the data set of $n=272$ eruptions which are contained in \cite{Hardle-book-1991} and also in \cite{Vena:Ripl:mode:1994}.
 Density estimation based on \added{same or different versions of} these data were also discussed by \cite{Silverman-1986-book}, \cite{Vena:Ripl:mode:1994}, and
 \cite{Leblanc-2010-JNS}.
\cite{Petrone-1999-CJS} provided  a comparison between the Baysian Bernstein and the kernel density estimates.

\begin{figure}
\begin{center}
  \makebox{\includegraphics[width=4.0in]{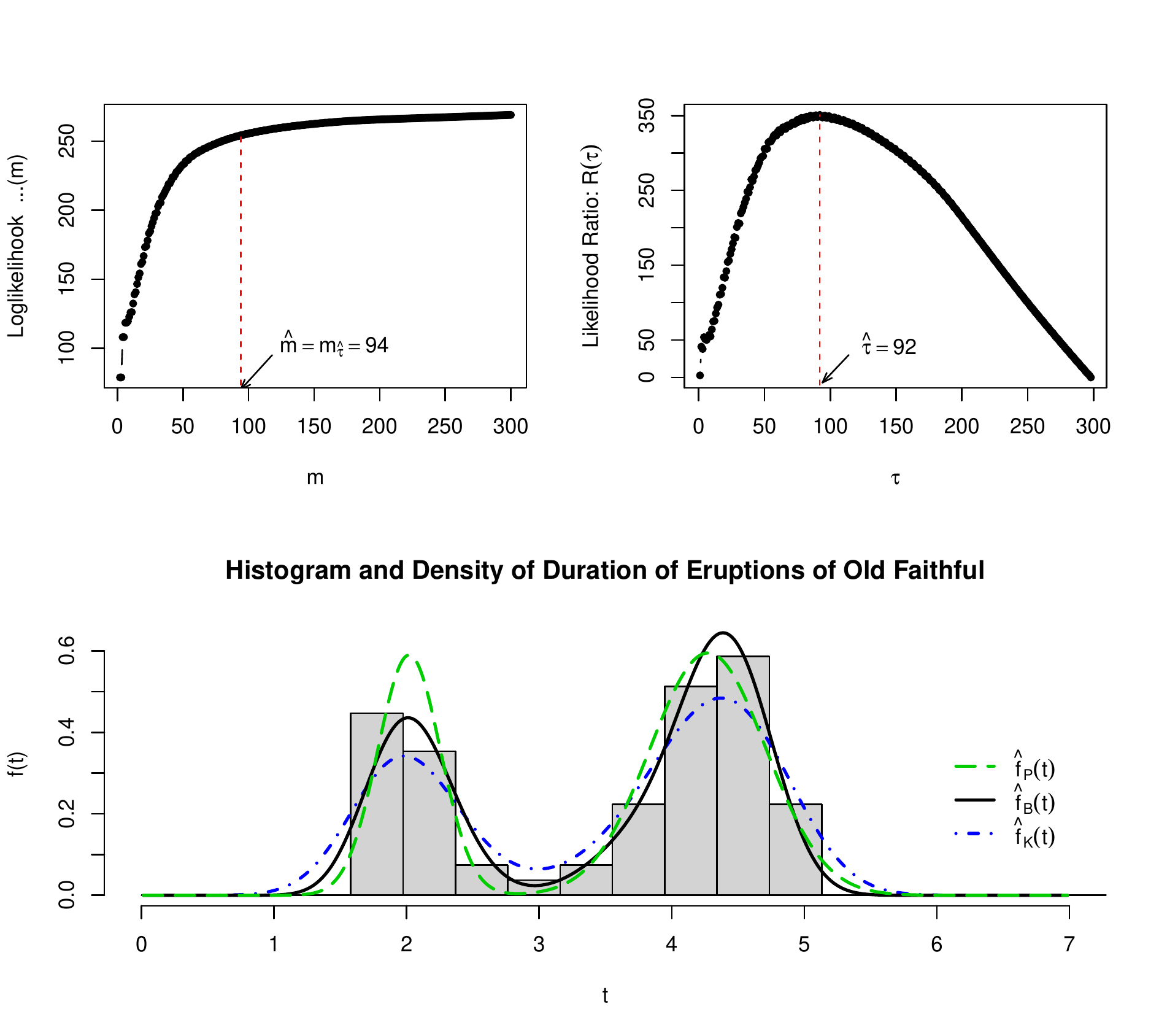}}
\caption{\label{fig: estimated-pdfs-based-on-Old-Faithful-data}
Upper panels: profile loglikelihood $\ell(m)$  and the likelihood ratio $R(\tau)$ for change-point $\tau$ for $m\in M\{2,3,\ldots,200\}$.
Lower panel: $\hat f_P$:
the parametric density estimate; $\hat f_K$: the kernel density estimate;  and $\hat f_B$: the proposed maximum  \bernstein likelihood estimate using
 $\hat m=94$.}
 \end{center}
\end{figure}

We truncate the data by interval $[a, b]=[0, 7]$ 
and transform the data to $y_i=x_i/b$, $i=1,\ldots,n$.
The maximum \bernstein likelihood estimate $\hat g_B$ of the truncated density based on the data $y_i$'s with the estimated optimal degree $\hat m=94$ is transformed to give the
maximum \bernstein likelihood estimate $\hat f_B$ of $f$:
 $\hat f_B(x)=\hat g_B\{(x-a)/(b-a)\}/(b-a).$
Figure \ref{fig: estimated-pdfs-based-on-Old-Faithful-data}   shows the histogram of the data and the estimated densities  using the  kernel method $\hat f_K$, the parametric method $\hat f_P$ using  R package \textsf{mixtools}'s function ``normalmixEM()'' which implements the  EM method for normal mixture model, and the proposed method $\hat f_B$ of this paper.

The difference between $\hat f_B$ and $\hat f_P$ could indicate that the normal mixture model does not  perfectly fit the data.
\subsection{A Microarray Data and the True Null Rate}
\begin{figure}
\begin{center}
  \makebox{\includegraphics[width=4.0in]{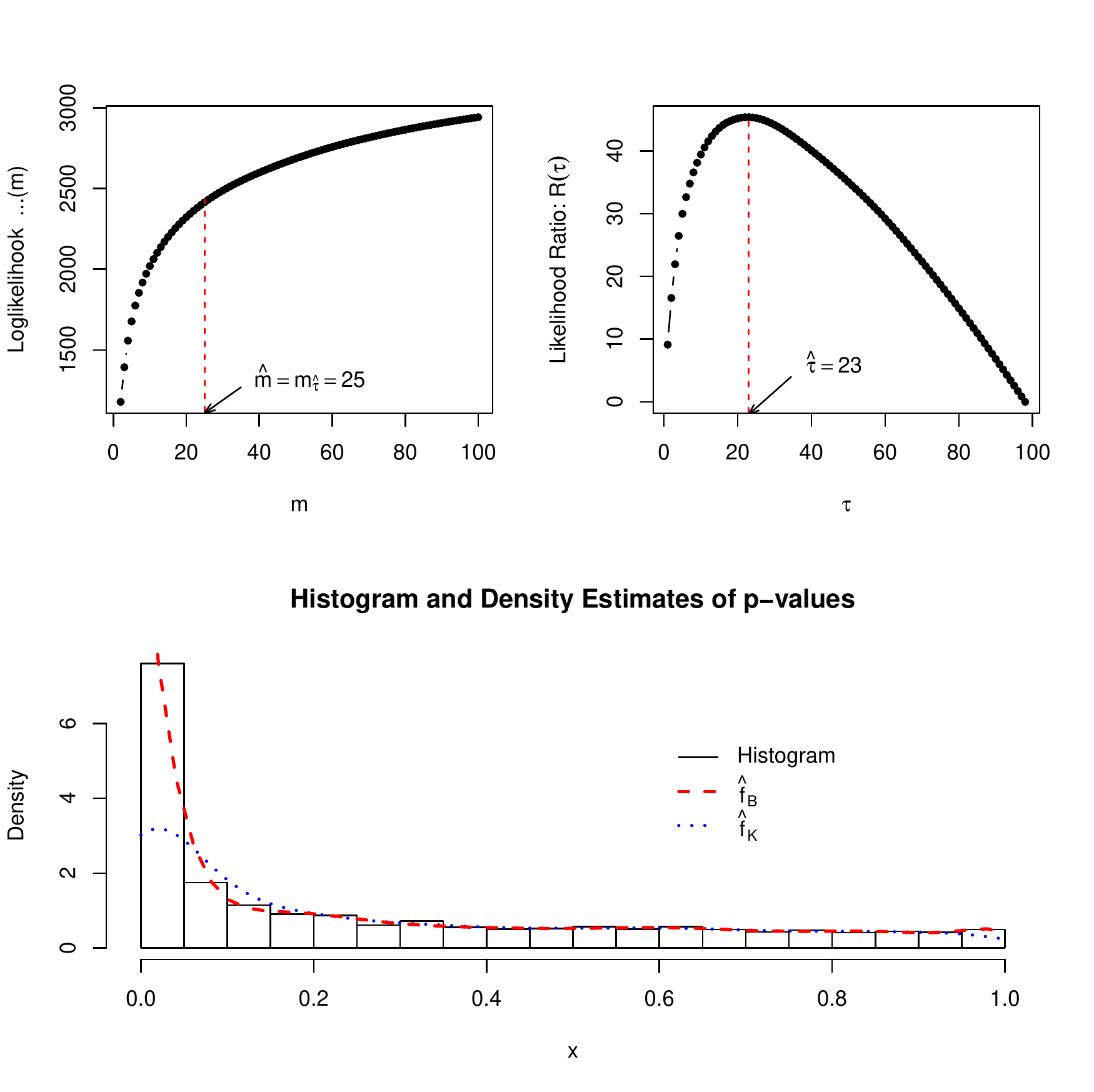}}
\caption{\label{fig: pval-density-optimal-m-25}
Upper panels: profile loglikelihood $\ell(m)$  and the likelihood ratio $R(\tau)$ for change-point $\tau$ for $m\in M\{2,3,\ldots,100\}$.
Lower panel: $\hat f_K$: the kernel density estimate;  and $\hat f_B$: the proposed maximum  \bernstein likelihood estimate using
 $\hat m=25$.}
 \end{center}
\end{figure}

When analyzing microarray data using multiple tests, the FDR is one of the important statistics \citep[see][for example]{Storey2002, Storey2003}. The key for calculating the FDR is the estimation of the proportion
$\pi_0$ of the true null hypotheses. Based on the $p$-values of the multiple tests, say $y_1,\ldots,y_n$, we first obtain an estimated density $\hat f$ of the $p$-values. The true density $f$ of the $p$-values is a mixture of uniform(0,1) density and another density $g$ on [0,1] so that $f(t)=\pi_0+(1-\pi_0)g(t)$, where $g$ is the density of $p$-values of the false null hypotheses. It is reasonable to assume that $g(1)=0$.  Then we can estimate $\pi_0$ by
$\hat \pi_0=\hat f(1)$ \citep[see][for example]{Guan:Wu:Zhao-2008}. The performance of this estimator relies on that of the density estimate $\hat f$ at the boundary $t=1$. The kernel estimate does not produce a good estimate due to its boundary effect. So \cite{Guan:Wu:Zhao-2008} proposed using $\tilde
f_B(t)$ for this purpose which has much less boundary effect. In Section 4 of \cite{Guan:Wu:Zhao-2008}, the leukemia gene expression dataset \citep{Golub1999} was considered. After data preprocessing and filtering, $p$-values of the $t$-test statistics were calculated by permutations for the remaining $n=3,571$ genes. They obtained the estimated true null rate $\tilde\pi_0=0.449$. Here we use the same set of $p$-values as an example of extremely skewed densities. We obtain $\hat m=25$ using the method of change-point. The estimated densities using the proposed  $\hat f_B$ and the kernel estimate $\hat f_K$ are shown in Figure \ref{fig: pval-density-optimal-m-25}. Clearly $\hat f_B$ performs much better than $\hat f_K$ especially at the boundaries.
 Using the proposed $\hat f_B$, we have $\hat\pi_{0B}=\hat f_B(1)=(\hat m+1)\hat p_{\hat m}=0.401$.

\section{Concluding Remarks} Unlike other applications of the \bernstein polynomial in statistics, in this paper we propose an approximate
parametric model for a nonparametric underlying density which enables people to use maximum likelihood method to solve the problem. One of the advantages of the proposed method over the others  such as the Bayesian method and even the kernel method is that there is only one tuning parameter $m$ to be determined to find a good model fit. Of course, the biggest advantage is that the proposed maximum \bernstein  likelihood density estimate achieves a  ``nearly parametric'' rate of convergence, \added{${\cal O}(\log n/n)$,} under some conditions. The proposed model is efficient because it is specific parametric one. Due to the ``global'' validity of the \bernstein approximation to a continuous function, the model shares the robustness of the nonparametric ones.  It is an interesting  project to generalize the proposed method to multivariate distributions.
The proposed model is not only for estimating a density function but  can also  be used  as a general nonparametric model to solve other statistical problems. As a nearly parametric solution of the nonparametric problem, the proposed
method not only has the robustness of nonparametric method but also has a nearly parametric efficiency.

Motivated by this approach and Theorem \ref{thm: rate of convergence of L2-distance btwn f and fB-hat}, we also see that other constructive approximations of functions on an interval finite or not
which have fast enough convergence
rate 
could also be used as an approximate parametric model for densities on the  interval.

\bibliographystyle{gNST}

\def\polhk#1{\setbox0=\hbox{#1}{\ooalign{\hidewidth
  \lower1.5ex\hbox{`}\hidewidth\crcr\unhbox0}}}
  \def\lfhook#1{\setbox0=\hbox{#1}{\ooalign{\hidewidth
  \lower1.5ex\hbox{'}\hidewidth\crcr\unhbox0}}} \def\cprime{$'$}
  \def\cprime{$'$}

\section*{Appendix: Proofs}
\subsection{Proof of Theorem \ref{thm: nested model}}
For each $f_B(t,\bm{p}_m)\in \mathscr{D}_m([0,1])$,
let $f_B(t,\bm{p}_m)=\sum_{i=0}^mp_{mi} \beta_{mi}(t)$ with $\sum_{i=0}^mp_{mi}=1$ and $p_{mi}\ge 0$.
By the binomial theorem,
\begin{eqnarray*}
  f_B(t,\bm{p}_m) &=& (m+1)\sum_{i=0}^mp_{mi} {m\choose i}t^i(1-t)^{m-i}\{t+(1-t)\}^r
  \\
   &=& (m+1)\sum_{i=0}^m\sum_{j=0}^r p_{mi} {m\choose i}{r\choose j}t^{i+j}(1-t)^{m+r-i-j}
  \\
   &=&  \sum_{j=0}^{m+r} p_{m+r,j} \beta_{m+r,j}(t),
\end{eqnarray*}
where $p_{m+r,j}=\frac{m+1}{m+r+1}\sum_{i=0}^m p_{mi} \frac{{m\choose i}{r\choose j-i}}{{m+r\choose j}}\ge 0$, $j=0,\ldots,m+r$.
So  $f_B(t,\bm{p}_m)\in \mathscr{D}_{m+r}([0,1])$.

\subsection{Proof of Theorem \ref{thm: rate of convergence of L2-distance btwn f and fB-hat}.}
Based on the assumptions of the theorem, we have
$$\int_0^1 \frac{\{\pi_{m}(t)-f(t)\}^4}{f^2(t)}I\{f(t)>0\}dt=\mathcal{O}(m^{-2k-\alpha}).$$
By Taylor expansion 
\begin{eqnarray*}\frac{1}{n}\{\ell(\pi_{m})-\ell(f)\}&=&\frac{1}{n}\sum_{j=1}^n\left\{\log \pi_{m}(x_j)- \log f(x_j)\right\}\\
&=&\frac{1}{n}\sum_{j=1}^n\frac{\pi_{m}(x_j)-f(x_j)}{f(x_j)}
-\frac{1}{2n}\sum_{j=1}^n \frac{\{\pi_{m}(x_j)-f(x_j)\}^2}{f^2(x_j)}+o(R_n),\; a.s.,
\end{eqnarray*}
where $R_n=C'm^{-2k}+C''_2m^{-2k}(\log\log n/n)^{1/2}
\le 
C'''n^{-2k/(2k+1)}\le C'''n^{-2/3}.$
By the law of iterated logarithm,
$$\frac{1}{n}\sum_{j=1}^n\frac{\pi_{m}(x_j)-f(x_j)}{f(x_j)}
=\mathcal{O}(n^{-5/6} \log\log n), \; a.s..$$
Thus we have
$$\frac{1}{n}\{\ell(\pi_{m})-\ell(f)\}
=-\frac{1}{2n}\sum_{j=1}^n \frac{\{\pi_{m}(x_j)-f(x_j)\}^2}{f^2(x_j)}+o(n^{-2/3}), \; a.s..$$
So $\hat f_B(t)=\beta_{m0}(t)+\hat{\bm{p}}^\tr\bar{\bm\beta}_{m}(t)$ satisfies
$$\frac{1}{n}\sum_{j=1}^n \frac{\{\hat f_{B}(x_j)-f(x_j)\}^2}{f^2(x_j)}\le \frac{1}{n}\sum_{j=1}^n \frac{\{\pi_{m}(x_j)-f(x_j)\}^2}{f^2(x_j)}=\mathcal{O}(n^{-2/3}), \; a.s..$$
It follows from this and (\ref{eq: approx of poly w pos coeff}) that
\begin{equation}\label{eq: weighted L2 distance btwn fb-hat and fmk}\frac{1}{n}\sum_{j=1}^n \frac{\{\hat f_{B}(x_j)-\pi_{m}(x_j)\}^2}{\pi^2_{m}(x_j)}
=\mathcal{O}(n^{-2/3}), \; a.s..
\end{equation}
Let ${\cal I}_{mn}={\cal I}_{mn}^{(r)}=\bigcup_{\hat{\bm p}}\{i\in (1,\ldots,m): |\hat p_i-p_i|>n^{-r}\}$.
It follows from this and  Taylor expansion at $\hat{\bm p}_{{\cal I}_{mn}}=\bm p_{{\cal I}_{mn}}$ that
$$0=\frac{\partial\ell_B(\hat{\bm{p}}_{{\cal I}_{mn}})}{\partial \bm{p}_{{\cal I}_{mn}}}
=\frac{\partial\ell_B(\bm{p}_{{\cal I}_{mn}})}{\partial \bm{p}_{{\cal I}_{mn}}}+
\frac{\partial^2\ell_B(\bm{p}_{{\cal I}_{mn}})}{\partial \bm{p}_{{\cal I}_{mn}}\partial \bm{p}_{{\cal I}_{mn}}^\tr}(\hat{\bm{p}}_{{\cal I}_{mn}}-\bm{p}_{{\cal I}_{mn}})+o(n^{-2/3}).$$
Thus we have
\begin{eqnarray*}
\hat f_B(t)-\pi_{m}(t)&=&\bar{\bm\beta}_{m}^\tr(t;{\cal I}_{mn})(\hat{\bm{p}}_{{\cal I}_{mn}}-\bm{p}_{{\cal I}_{mn}})+\bar{\bm\beta}_{m}^\tr(t;{\cal I}_{mn}^c)(\hat{\bm{p}}_{{\cal I}_{mn}^c}-{\bm{p}}_{{\cal I}_{mn}^c}) \\
&=&\bar{\bm\beta}_{mn}^\tr(t;{\cal I}_{mn})\left[\frac{\partial^2\ell_B(\bm{p}_{{\cal I}_{mn}})}{\partial \bm{p}_{{\cal I}_{mn}}\partial \bm{p}_{{\cal I}_{mn}}^\tr}\right]^{-1}
\left\{-\frac{\partial\ell_B(\bm{p}_{{\cal I}_{mn}})}{\partial \bm{p}_{{\cal I}_{mn}}}+ o(n^{1/3})\right\}+{\cal O}(mn^{-r})\\
&=&\bar{\bm\beta}_{m}^\tr(t;{\cal I}_{mn})\left[I_m(\pi_{m}; {\cal I}_{mn})\right]^{-1}
 \frac{1}{n}\sum_{j=1}^n\frac{\bar{\bm\beta}_{m}(x_j;{\cal I}_{mn})}{\pi_{m}(x_j)}+ o(n^{-2/3})+{\cal O}(mn^{-r}).
\end{eqnarray*}
Because $m\le C_2n^{1/2k}$, $mn^{-r}\le C_2n^{1/2k-r}$ and $r>(k+1)/2k$, $mn^{-r}=o(n^{-1/2})$.
So $\sqrt{n}\{\hat f_B(t)-\pi_{m}(t)\}/\sigma_{mk}(t)$ converges to normal $N(0,1)$ in distribution, where
\begin{eqnarray*}
\sigma_{mk}^2(t)&=&\Var\{\hat f_B(t)-\pi_{m}(t)\}\\
&=&\bar{\bm\beta}_{m}^\tr(t;{\cal I}_{mn})\left[I_m(\pi_{m}; {\cal I}_{mn})\right]^{-1}\bar{\bm\beta}_{m}(t;{\cal I}_{mn})+o(n^{-1})\\
&\le& C_4\bar{\bm\beta}_{m}^\tr(t;{\cal I}_{mn})\left[I_m(1; {\cal I}_{mn})\right]^{-1}\bar{\bm\beta}_{m}(t;{\cal I}_{mn})+o(n^{-1}).
\end{eqnarray*}
Integrating $\sigma_{mk}^2(t)$ we get
\begin{eqnarray*}
\varsigma^2_{m}(\pi_{m})&=&\int_0^1\sigma^2_{mk}(t)dt \\
&\le&C_4 \mathrm{trace}\left\{\left[I_m(1; {\cal I}_{mn})\right]^{-1}\int_0^1\bar{\bm\beta}_{m}(t;{\cal I}_{mn})\bar{\bm\beta}_{m}^\tr(t;{\cal I}_{mn})dt\right\}+o(n^{-1})
\\
&=&C_4 \mathrm{trace}\{\left[I_m(1; {\cal I}_{mn})\right]^{-1}I_m(1; {\cal I}_{mn})\}+o(n^{-1})\\
&\le& C_4 \kappa_{mn}^{(r)}+o(n^{-1}).
\end{eqnarray*}
 Therefore $E\int_0^1\{\hat f_B(t)-\pi_{m}(t)\}^2dt = \varsigma^2_{m}(\pi_{m}) \mathcal{O}(n^{-1})
=\kappa_{mn}^{(r)} \mathcal{O}(n^{-1})
$ and
\begin{eqnarray*}
E\int_0^1\{\hat f_B(t)-f(t)\}^2dt&\le& 2E\int_0^1\{\hat f_B(t)-\pi_{m}(t)\}^2dt+
2\int_0^1\{f(t)-\pi_{m}(t)\}^2dt\\
&=& \mathcal{O}(\kappa_{mn}^{(r)} n^{-1})+\mathcal{O}(m^{-2k-\alpha}).
\end{eqnarray*}
The proof of Theorem \ref{thm: rate of convergence of L2-distance btwn f and fB-hat} is complete.

\end{document}